\documentclass{an}
\usepackage{graphicx}
\usepackage{times}
\usepackage{fancyhdr}
\begin{document}

\title{GQ\,Lup and its common proper motion companion}

\author{M. Mugrauer \inst{1} and R. Neuh\"auser \inst{1}}
\institute{Astrophysikalisches Institut und Universit\"ats-Sternwarte, Schillerg\"a{\ss}chen 2,
D-07745 Jena, Germany}

\date{Received; accepted; published online}

\abstract{Recently, Neuh\"auser et al. (2005a) presented evidence for a sub-stellar, common proper
motion companion to GQ\,Lup. With two theoretical mass estimates, both below the Deuterium burning
minimum mass limit, the companion is probably a planet imaged directly. We present here a more
detailed astrometric analysis of the GQ\,Lup system, using all the (different) proper motions
published for the primary. The common proper motion is significant in all cases, also when taking
into account the error in parallax or distance ($\rm{140\pm50}$\,pc). When using the weighted mean,
the significance for common proper motion of GQ\,Lup and its companion is 7\,$\sigma$ + 4\,$\sigma$
for no change in separation plus 8\,$\sigma$ for no change in position angle. We also discuss the
question, whether GQ\,Lup and its common-proper motion companion are not bound, but share the same
or similar proper motion as two independent members of the Lupus T association, which is a moving
group, where most members should have the same motion anyway. Given our discussion, this hypothesis
can be rejected by several $\sigma$: The probability to find by chance an L-dwarf fainter than
$\rm{K_{S}}$\,=\,14\,mag within 0.7325\,$^{\prime\prime}$ with (almost) the same proper motion of
GQ\,Lup is only $\rm{\le 3 \cdot 10^{-10}}$. The orbital motion of the system is not yet detected
($\rm{1.4\pm2.2}$\,mas/yr), but is probably smaller than the escape velocity
($\rm{5.3\pm2.1}$\,mas/yr), so that the system may well be gravitationally bound and stable. This
is different for the 2MASSWJ\,1207334-393254 system, as we also show. \keywords{GQ Lup A, GQ Lup b,
2M1207, extra-solar planets, late-type stars, astrometry}}

\correspondence{markus@astro.uni-jena.de}

\maketitle

\section{Introduction: GQ\,Lup and its companion}

\begin{figure}
\resizebox{\hsize}{!}{\includegraphics{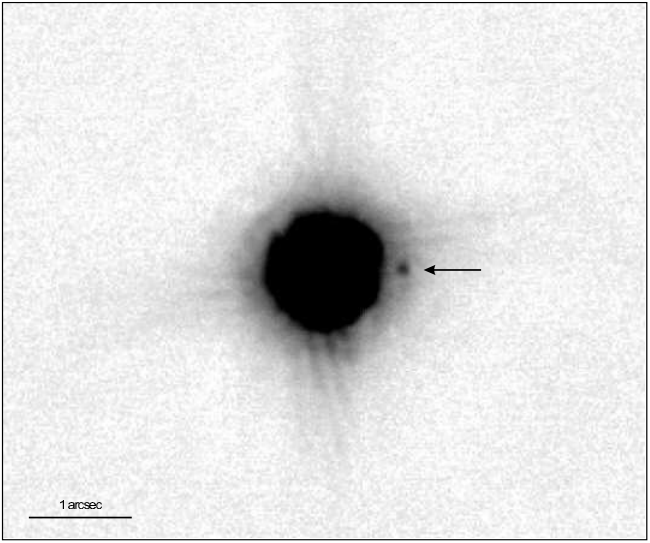}} \caption{Subaru image of GQ\,Lup\,A and
companion b in the L-band.}\label{subaru}
\end{figure}

\begin{figure}
\resizebox{\hsize}{!}{\includegraphics{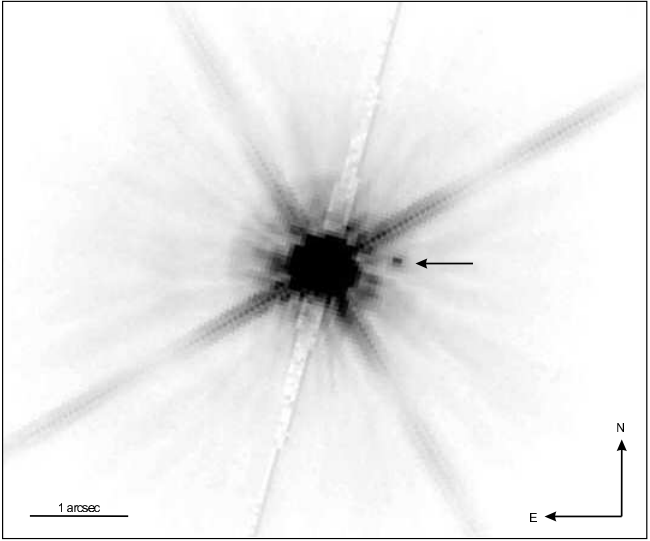}} \caption{HST/PC image of GQ\,Lup\,A and companion
b in filter F606W.}\label{hst}
\end{figure}

With $\rm{K_{S}}$ - band imaging using our own imaging data obtained with the Very Large Telescope
(VLT) and its infrared adaptive optics camera NAos-COnica (NACO), we detected a $\sim$\,6\,mag
fainter object $\rm{0.7325\pm0.0034}$\,$^{\prime\prime}$ west of the classical T Tauri star
GQ\,Lup, located in the Lupus I cloud. Using archival data from the Subaru telescope adaptive
optics camera (CIAO), see Fig.\,\ref{subaru}, and the Hubble Space Telescope (HST) Planetary Camera
(PC), see Fig.\,\ref{hst}, we could show that it is clearly a co-moving companion (Neuh\"auser et
al. 2005a, henceforth N05a), but orbital motion was not yet detectable.

The NACO $K\rm{_{S}}$ - band spectrum yielded $\sim$\,L1-2 (M9-L4) as spectral type. At
$\rm{140\pm50}$\,pc distance (Lupus I cloud), it can be placed into the H-R diagram. According to
our own calculations following Wuchterl \& Tscharnuter (2003), it has 1 to 3\,$\rm{M_{Jup}}$. This
model takes into account the formation and collapse. A comparison of our spectrum with the
GAIA-dusty model (Brott \& Hauschildt, in preparation), an update of the AMES-dusty model (Allard
et al. 2001), yielded the radius to be $\rm{1.2\pm0.5}$\,$\rm{R_{Jup}}$ and the gravity to be
$\rm{\log g}$\,=\,2.0 to 3.3 (g in cgs units). For M9-L4 spectral type, the temperature is
$\rm{2050\pm450}$\,K. At the given flux of the GQ\,Lup companion ($\rm{K_{S}=13.10\pm0.15}$\,mag at
$\rm{140\pm50}$\,pc), this yields a mass of $\rm{\le2}$\,$\rm{M_{Jup}}$ (for $\rm{\log g
\simeq}$\,4 and 2\,$\rm{R_{Jup}}$, it is $\sim$\,6\,$\rm{M_{Jup}}$), see also Neuh\"auser et al.
(2005b), henceforth N05b.

Mohanty et al. (2004a) measured gravities for isolated young brown dwarfs and free-floating
planetary mass objects. Their coolest objects have spectral type M7.5 and gravities as low as
$\rm{\log g=3.125}$ (GG\,Tau\,Bb). This lead Mohanty et al. (2004b) to mass estimations as low as
$\sim$\,10\,$\rm{M_{Jup}}$. GQ\,Lup\,A is younger than the Mohanty et al. Upper Sco objects. Its
companion is at least as late in spectral type, probably even cooler. An object younger and cooler
must be lower in mass. The GQ\,Lup companion is fainter than the faintest Mohanty et al. object
(USco\,128, $\sim$\,9\,$\rm{M_{Jup}}$), so that the mass estimate for the GQ\,Lup companion would
be $\rm{\le8}$\,$\rm{M_{Jup}}$ (see also N05b). However, Reiners\footnote{see
http://www.iac.es/workshop/ulmsf05/pres/reiners.pdf} argues that some TiO oscillator strength were
wrong in some inputs used in Mohanty et al. (2004a,b), so that all temperatures have to be
increased by 150 to 200\,K, and, hence, also mass estimates. Therefore, the mass estimate for
USco\,128 and, hence, the upper mass limit for GQ\,Lup\,b is now $\sim$\,15\,$\rm{M_{Jup}}$.

According to Burrows et al. (1997) and Baraffe et al. (2002), the mass of the companion could be
anywhere between 3 and 42\,$\rm{M_{Jup}}$, as given in N05a. However, these models are not valid at
the young age of GQ\,Lup, so that they are not applicable. According to both Mohanty et al. (2004b)
and Close et al. (2005), the Baraffe et al. (2002) models overestimate the masses of young
sub-stellar objects below $\sim$\,30\,$\rm{M_{Jup}}$, but may underestimate them above
$\sim$\,40\,$\rm{M_{Jup}}$. This is consistent with our results.

Hence, according to all valid estimations, the mass of the GQ\,Lup companion is almost certainly
below $\sim$\,13\,$\rm{M_{Jup}}$, hence probably a planet imaged directly, to be called GQ\,Lup\,b.
For more details, see N05a and N05b.

We present all the different proper motions published for GQ\,Lup in section 2 and discuss in
section 3 the question, whether the primary star and its faint companion candidate form a common
proper motion pair according to all the published proper motions or their weighted mean, also
taking into account the error in parallax of GQ\,Lup. Then, in section 4, we discuss an alternative
interpretation, namely that GQ\,Lup and the faint object, even though sharing the same proper
motion and being located within 1\,$^{\prime\prime}$, are not bound, but at slightly different
distances, both within the Lupus I cloud. We also compare the orbital properties of the common
proper motion companions of GQ\,Lup and the brown dwarf 2MASSWJ\,1207334-393254 (afterwards
2M1207). We conclude in section 5.

\section{The proper motion of GQ\,Lup}

To investigate, whether a star and its companion candidate form a common proper motion pair, one
has to check, whether the separation and position angle (PA) between the two is constant over time
(however, one has to allow for a small change in separation and/or PA because of orbital motion).
The epoch difference needed for significant results is given by the proper (and parallactic) motion
of the star and the precision achieved in the imaging. The background hypothesis is that the
companion candidate has negligible proper and parallactic motion -- negligible compared to the
primary star.

\begin{table} [tb]
\caption{Proper motions of GQ\,Lup and the estimated weighted mean}
\begin{tabular}{lcc}
Reference  & $\rm{\mu_{\alpha}cos(\delta)}$ & $\rm{\mu_{\delta}}$ \\
\hline & [mas/yr] & [mas/yr]\\ \hline
Carlsberg Meridian Cat. 1999 & -20.2$\pm$3.9 & -21.8$\pm$4.1 \\
Kharchenko 2001              & -20.55$\pm$3.89 & -22.1$\pm$4.0\\
USNO-B1.0 (Monet et al. 2003)&  -11.4$\pm$14.6 & -50$\pm$10\\
UCAC2 (Zacharias et al. 2004)& -18.2$\pm$4.6   & -22.2$\pm$4.7\\
Camargo et al. 2003          & -14$\pm$3       & -21$\pm$3\\
Teixeira et al. 2000         & -27$\pm$4       & -14$\pm$4\\
\hline weighted mean & -19.15$\pm$1.67 & -21.06$\pm$1.69\\ \hline
\end{tabular}\label{pm}
\end{table}

\begin{figure}
\resizebox{\hsize}{!}{\includegraphics{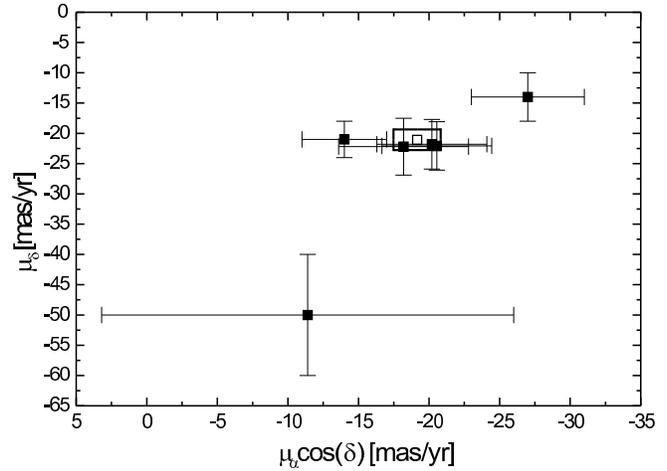}} \caption{Proper motions of GQ\,Lup (see
Table\,\ref{pm}). The weighted mean is shown in the center (open square) with a 1\,$\sigma$ error
box.}\label{pmplot}
\end{figure}

There are six different measurements of the proper motion for GQ\,Lup given in the literature, see
Table\,\ref{pm} and Fig.\ref{pmplot}. Five of them have a typical precision of 3 to\,5 mas/yr. Only
the USNO data are an exception with 10 to 15\,mas/yr precision. In N05a, we list the two extremes
of the remaining five proper motions, namely Teixeira et al. (2000) and Camargo et al. (2003),
which are also the best in precision (3 to 4\,mas/yr). In Table\,\ref{pm}, we also list the
computed weighted mean of the six measurements.

\section{The common proper motion pair}

Here, we present the estimated change in both separation and position angle (PA) for all different
proper motions published as well as for the weighted mean -- always for two alternatives: Either
the two objects form a common-proper motion pair (i.e. separation and PA are constant $\pm$ orbital
motion), or the faint object is a non-moving background object with negligible proper and
parallactic motion, so that only the star moves resulting in a change in separation and PA.

We compare the estimated changes with the actually observed data for GQ\,Lup in
Table\,\ref{astroall} (observed astrometric data repeated here in Table\,\ref{astrometry} from N05a
for convenience). We then always estimate the significance, by which we can reject the background
hypothesis, see Table\,\ref{parallax} and Table\ref{astroall}.

\begin{table}
\caption{Astrometry of the GQ\,Lup system (Neuh\"auser et al. 2005a)}
\begin{tabular}{lcccc}
Telescope & Instr. & Epoch & Separation & PA\\
\hline & & mm/yyyy & [mas] &  [$^{\circ}$]\\ \hline
HST       & PC   & 04/1999 & 739$\pm$11 & 275.62$\pm$0.86\\
Subaru    & CIAO & 07/2002 & 736.5$\pm$5.7 & (*) \\
VLT       & NACO & 06/2004 & 732.5$\pm$3.4 & 275.45$\pm$0.3 \\
VLT       & NACO & 08/2004 & 731.4$\pm$4.2 & (*)\\
VLT       & NACO & 09/2004 & 735.8$\pm$3.7 & (*)\\ \hline
\end{tabular}\label{astrometry}
(*) No astrometric calibration available for detector orientation, see N05a.
\end{table}

Let us first consider the possible effect of parallax uncertainty on the significance of common
proper motion. Prior to the Hipparcos mission, the distance of $\sim$\,140\,pc was usually used for
the Lupus star forming clouds including the Lupus I cloud with GQ\,Lup (see e.g. Krautter 1991 and
Hughes et al. 1993 for a discussion). The Hipparcos measurements of the T Tauri stars in Lupus
confirms this result with $\rm{145\pm50}$\,pc mean distance (see Wichmann et al. 1998 for T Tauri
stars known before the ROSAT mission and Neuh\"auser \& Brandner 1998 for T Tauri stars newly found
by the ROSAT mission). However, some members of Lupus also appeared to be as far as $\sim$\,190\,pc
(Wichmann et al. 1998), while Knude \& Hog (1998) estimated $\sim$\,100\,pc for the distance of
Lupus I. More recently, de Zeeuw et al. (1999) list $\rm{142\pm2}$\,pc for Upper Centaurus and
Lupus; Nakajima et al. (2000) give $\sim\,150$\,pc for Lupus; Franco (2002) argue for
$\sim\,150$\,pc {\em as the most probable distance to the dark cloud known as Lupus I}; and Sartori
et al. (2003) specify 147\,pc. Hence, we use $\rm{140\pm50}$\,pc for the distance towards GQ\,Lup
in both N05a and here. This results in a parallax of 5.3 to 11.1\,mas (7.1\,mas for 140\,pc).

\begin{table*}
\caption{Comparison of significance for common proper motion for the whole range of parallaxes
possible (by using the weighted mean for the proper motion of GQ\,Lup\,A): We list all measured
separations and position angles (PA) and the estimated separations for the background hypothesis
(assuming that the companion does not move). The significance of rejecting the background hypothesis
is also given. The 2004 epoch is 25 June 2004, our deepest NACO image. No PA is available for
Subaru 2002, because those data were taken by us from the public archive without calibration
available.} \centering{
\begin{tabular}[htb]{lccccccc}
Parallax  & Epoch     & \multicolumn{3}{c}{Separation [mas]} & \multicolumn{3}{c}{PA [$^{\circ}$]} \\
          &           & predicted (*) & observed & $\sigma$ & predicted (*) & observed & $\sigma$ \\ \hline
5.3\,mas  & 1999-2004 & 836.2$\pm$9.4 & 739$\pm$11    & 6.7 & 267.3$\pm$0.6 & 275.62$\pm$0.86 & 7.9 \\
          & 2002-2004 & 765.6$\pm$4.7  & 736.5$\pm$5.7 & 3.9 & & & \\ \hline
7.1\,mas  & 1999-2004 & 838.3$\pm$9.4 & 739$\pm$11    & 6.9 & 267.4$\pm$0.6 & 275.62$\pm$0.86 & 7.8 \\
          & 2002-2004 & 765.1$\pm$4.7  & 736.5$\pm$5.7 & 3.9 & & & \\ \hline
11.1\,mas & 1999-2004 & 843.0$\pm$9.4 & 739$\pm$11    & 7.2 & 267.4$\pm$0.6 & 275.62$\pm$0.86 & 7.8 \\
          & 2002-2004 & 764.1$\pm$4.7  & 736.5$\pm$5.7 & 3.7 & & & \\ \hline
\end{tabular}

Remark: (*) estimated for the background hypothesis, i.e. if not co-moving, which can be rejected.

}\label{parallax}

\end{table*}

\begin{figure}[htb]
\resizebox{\hsize}{!}{\includegraphics{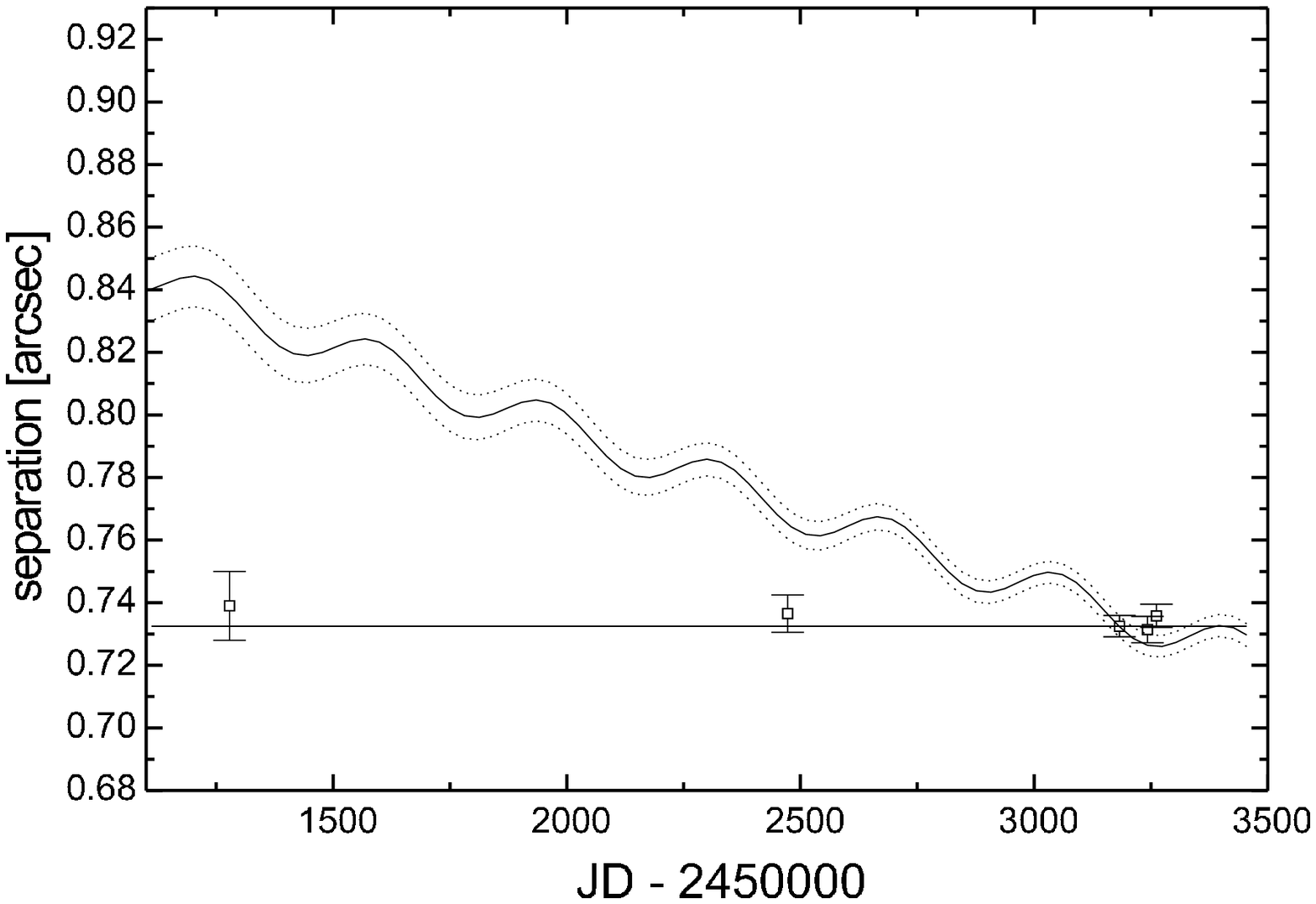}}\linebreak\linebreak\linebreak
\resizebox{\hsize}{!}{\includegraphics{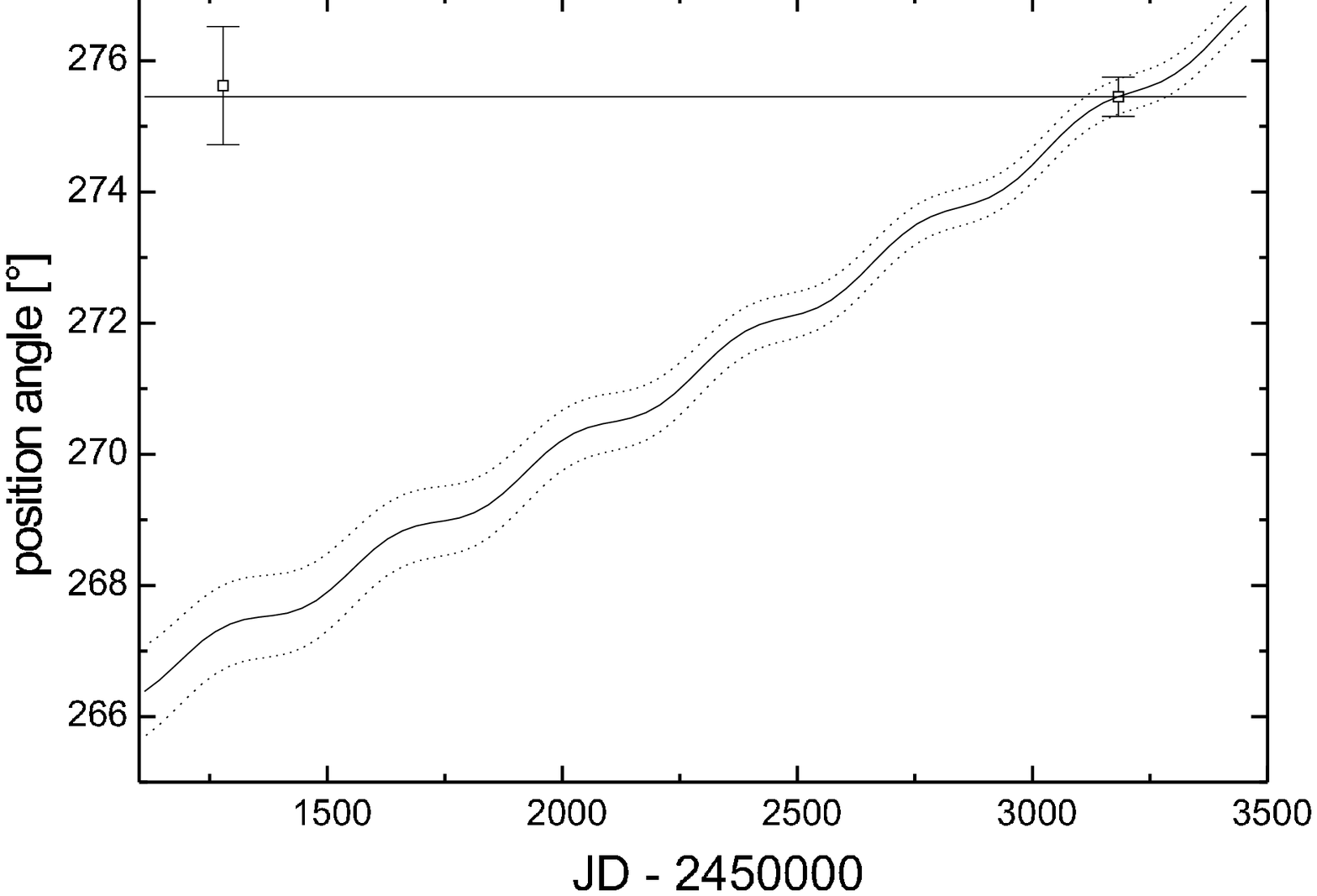}} \caption{Change in separation (top) and
position angle (bottom) according to weighted mean proper motion (curved full lines) for background
hypothesis with uncertainties from errors in proper motion (curved dotted line), which can be
rejected; no change in separation (left) or PA (right) as full straight lines with data points from
Table\,\ref{astroall}.}
\end{figure}

In Fig.\,\ref{astroall}, we show the expected change in separation and PA compared to the
observations, estimated for the mean proper motion of GQ\,Lup as given in Table\,\ref{pm}.
Table\,\ref{parallax} lists the significance, by which the background hypothesis can be rejected
for the three different parallaxes, for both change in separation and PA for two independent
different epoch differences available (1999-to-2004 comparing HST to VLT and 2002-to-2004 comparing
Subaru to VLT; for the 2004 epoch, we use our deepest NACO image from 25 June 2004).

\begin{figure*} [ht]
\resizebox{\hsize}{!}{\includegraphics[height=5cm]{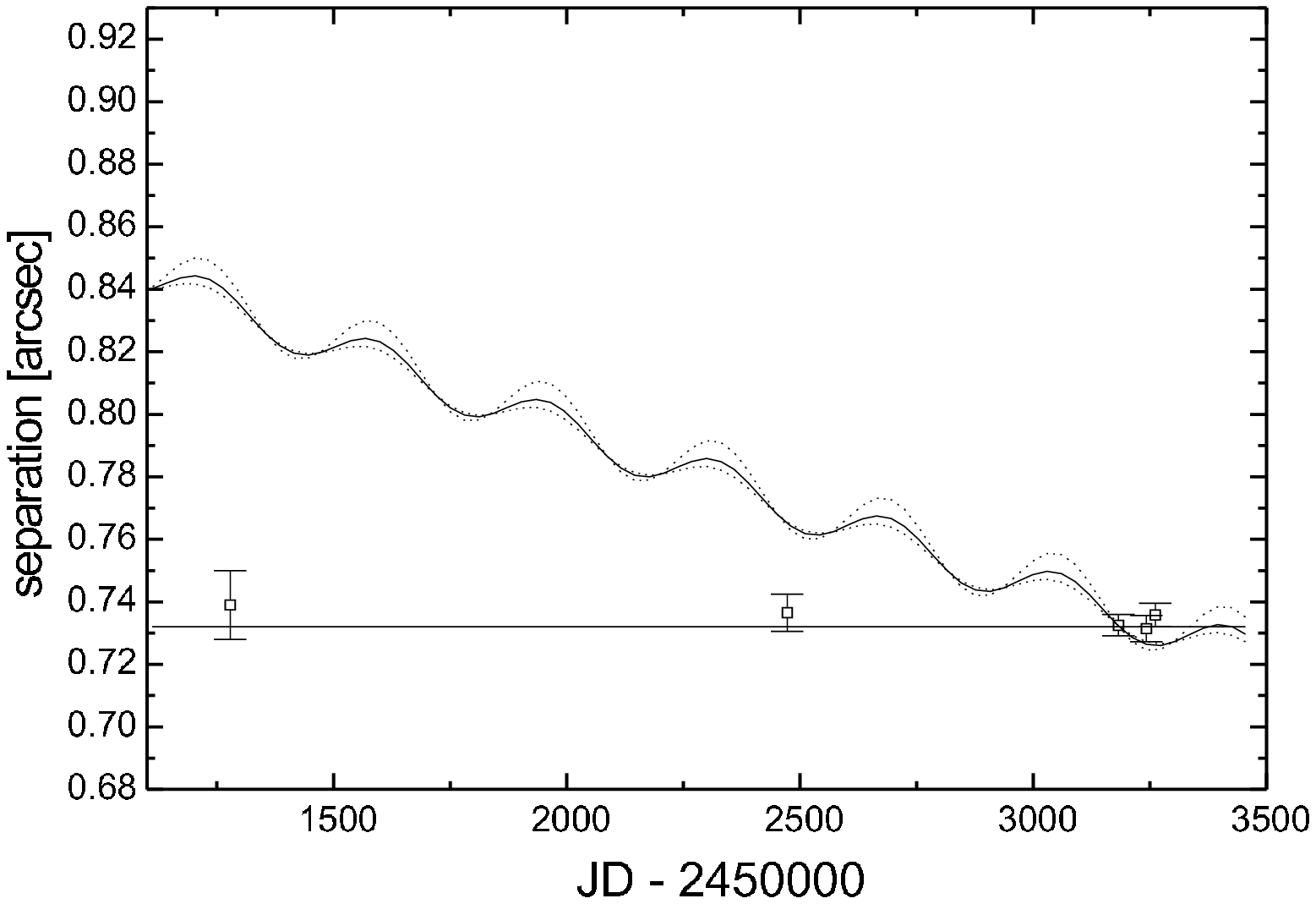}
\includegraphics[height=4.8cm]{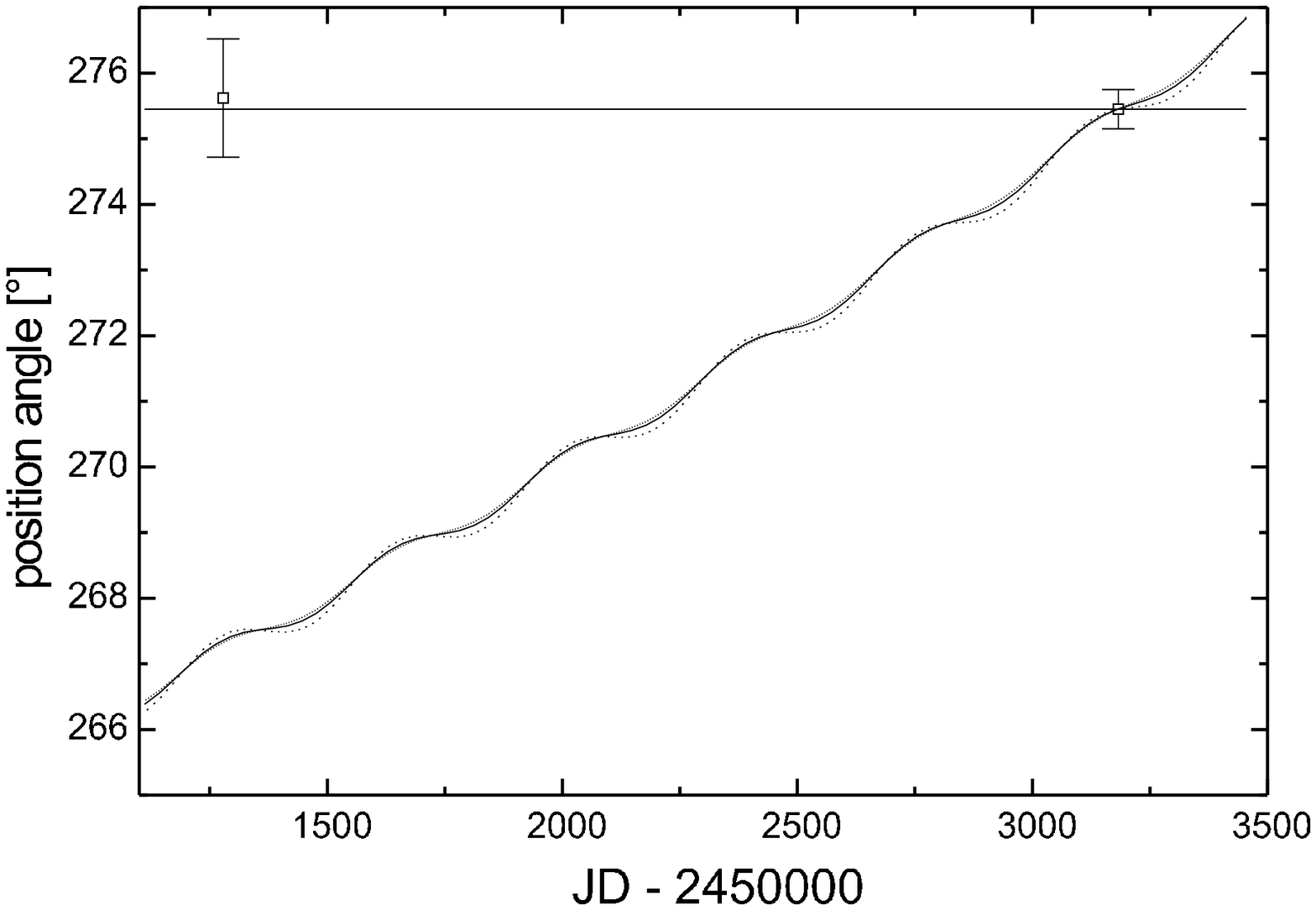}} \caption{The effect of the parallax uncertainty on the
separation (left) and the position angle PA (right). We plot the data points from HST, Subaru and
VLT (no position angle for Subaru, because those data were taken from the archive without
calibration available to us) versus observing epoch. The full straight line indicates no change in
separation. The curved full line is the expected separation for a parallax of 7.1\,mas, the dotted
lines for 5.3\,mas and 11.1\,mas (i.e. $\rm{140\pm50}$\,pc). The mean proper motion is used for
GQ\,Lup\,A. The background hypothesis can be rejected by $\rm{\ge10\sigma}$ (see
Table\,\ref{parallax}).}\label{par_seppa}
\end{figure*}

As can be seen in Table\,\ref{parallax}, Fig.\,\ref{par_seppa}, the effect of parallax uncertainty
is very small, only $\rm{\pm0.3}$\,dex in the significance in separation, and vanishing in
significance in PA (only $\rm{\pm0.1}$\,dex), hence negligible (and actually neglected in N05a). In
the remainder of this paper (Table\,\ref{astroall}), we have used 7.1\,mas as parallax (140\,pc).

As can be seen from Table\,\ref{astroall}, there is always high significance for common proper
motion. Even in the worst case of the USNO proper motion, which is far off the others and the
weighted mean and which has the largest errors, the significance for common proper motion is still
above 4\,$\sigma$ by rejecting the background hypothesis in PA by 4.2\,$\sigma$. For the weighted
mean proper motion, the significance for common proper motion is 6.9\,$\sigma$ by comparing the
1999 to 2004 separations, plus 3.9\,$\sigma$ by comparing the 2002 to 2004 separations, plus
7.8\,$\sigma$ by comparing the 1999 to 2004 PAs. These significances can be added up properly. In
total, its well above 10\,$\sigma$ (and, hence, even higher than given in N05a using the Teixeira
et al. (2000) proper motion only). Hence, GQ\,Lup\,A and the faint object west of it clearly form a
common proper motion pair.

\begin{table*}[htb]
\caption{List of all estimated and observed separations and positions angles (PA), the significance
to reject the background hypothesis is also given. The 2004 epoch is 25 June 2004, our deepest NACO
image. No PA is available for Subaru 2002, because those data were taken by us from the public
archive without calibration available.} \centering{
\begin{tabular}{lccccccc}
PM         & Epoch      & \multicolumn{3}{c}{Separation [mas]} & \multicolumn{3}{c}{PA [$^{\circ}$]} \\
reference  & difference & predicted (*) & observed & $\sigma$ & predicted (*) & observed & $\sigma$ \\
\hline
Carlsberg  & 1999-2004 & 844.0$\pm$21.1 & 739$\pm$11 & 4.4 & 267.1$\pm$1.4 & 275.62$\pm$0.86 & 5.2  \\
           & 2002-2004 & 767.1$\pm$8.5 & 736.5$\pm$5.7 & 3.0 \\ \hline
Kharchenko & 1999-2004 & 845.9$\pm$ 21.1 & 739$\pm$11 & 4.5 & 267.0$\pm$1.4 & 275.62$\pm$0.86 & 5.3 \\
           & 2002-2004 & 767.8$\pm$ 8.5 & 736.5$\pm$5.7 & 3.1\\ \hline
USNO       & 1999-2004 & 819.2$\pm$64.2 & 739$\pm$11 & 1.2 & 256.6$\pm$4.5 & 275.62$\pm$0.86 & 4.2 \\
           & 2002-2004 & 750.1$\pm$24.1 & 736.5$\pm$5.7 & 0.6\\ \hline
UCAC2      & 1999-2004 & 833.7$\pm$24.7 & 739$\pm$11 & 3.5 & 266.9$\pm$1.7 & 275.62$\pm$0.86 & 4.6 \\
           & 2002-2004 & 763.2$\pm$9.7 & 736.5$\pm$5.7 & 2.4 \\ \hline
Camargo    & 1999-2004 & 811.5$\pm$16.0 & 739$\pm$11 & 3.7 & 267.3$\pm$1.1 & 275.62$\pm$0.86 & 6.0  \\
           & 2002-2004 & 755.1$\pm$6.8 & 736.5$\pm$5.7 & 2.1 \\ \hline
Teixeira   & 1999-2004 & 878.4$\pm$21.1 & 739$\pm$11 & 5.9 & 269.9$\pm$1.4 & 275.62$\pm$0.86 & 3.5   \\
           & 2002-2004 & 780.9$\pm$8.5 & 736.5$\pm$5.7 & 4.3 \\ \hline
weigthed   & 1999-2004 & 838.3$\pm$9.4 & 739$\pm$11 & 6.9 & 267.4$\pm$ 0.6 & 275.62$\pm$0.86 & 7.8  \\
mean       & 2002-2004 & 765.1$\pm$4.7 & 736.5$\pm$5.7 & 3.9 \\ \hline
\end{tabular}

Remark: (*) estimated for background hypothesis, i.e. if not co-moving, which can be rejected.

}\label{astroall}

\end{table*}

\section{An alternative Interpretation}

In principle, the two alternatives studied, namely that the companion (candidate) either has
exactly the same proper motion as GQ\,Lup\,A or has neither proper nor parallactic motion at all,
are unrealistic extremes. Even if the star and the faint object form a common proper motion pair as
being gravitationally bound, one has to expect orbital motion, so that separation and/or PA change
slowly but periodically (see below). And even if the two objects are not gravitationally bound,
the fainter one does not necessarily have to stand still, it will have some proper and parallactic
motion. If it is a background object, though, its proper and parallactic motion should be much
smaller than for the primary star. But if it would be a foreground object,
its motion could be larger than for GQ\,Lup\,A.

It is, however, very unlikely that two objects have exactly the same proper motion, but different
distances: Our L1-2 companion could be either a $\sim$\,1\,Myr planet of GQ\,Lup or a Gyrs old L1-2
dwarf at $\sim$\,26\,pc ($\rm{M_{K}}$\,=\,11\,mag for L1-2, Tinney et al. 2003) with a space motion
as high as to mimic the same proper motion as GQ\,Lup, or an intermediate-age object between 26 and
140\,pc. With 100 field L-dwarfs brighter than K\,=\,14\,mag on the whole sky
(spider.ipac.caltech.edu/staff/davy/ARCHIVE), the chance to find one within a circle with
0.7325\,$^{\prime\prime}$ radius is $\rm{\le3\cdot10^{-10}}$. The probability to find one such
object also with the same proper motion is even smaller. The number is almost the same when
considering the whole spectral type range of M9 to L4 given in N05a. Given $\rm{0.0057\pm0.0025}$
L-dwarfs per $pc^{3}$ (Gizis et al. 2001), we expect only $\sim$\,$\rm{2\times10^{-7}}$ (or
$\sim$\,$\rm{5\times10^{-7}}$, respectively) L-dwarfs in the frustum of a right circular cone
between 26 and 140\,pc (or even 190\,pc, respectively), within a circle with
0.7325\,$^{\prime\prime}$ radius. It is even less likely to find one within this circle or volume
having even the same proper motion as GQ\,Lup\,A.

\begin{figure}[h]
\resizebox{\hsize}{!}{\includegraphics{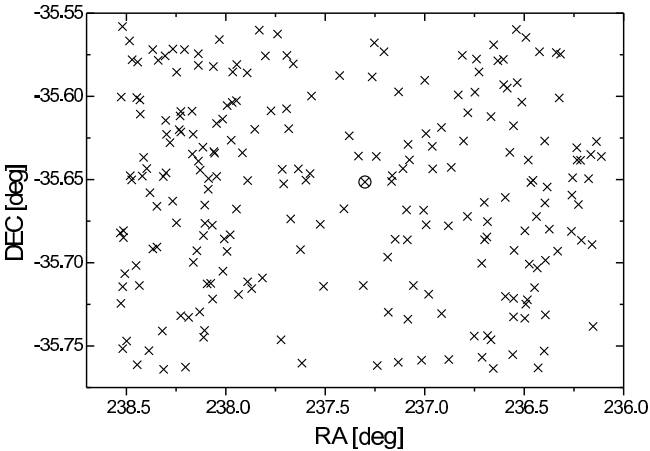}} \caption{Positions of all stars listed in
Camargo et al. (2003) within 0.4 square degree around GQ\,Lup (indicated in the
center).}\label{poscamargo}
\end{figure}

\subsection{Same proper motion, both in Lupus, but not bound ?}

As discussed in Neuh\"auser et al. (2002) for a late-M-type primary with a nearby fainter companion
candidate in the Chamaeleon I cloud, which was later found to be a background giant (Neuh\"auser et
al. 2003), one has to take into account that most objects in a T association (like Chamaeleon and
Lupus) do have a similar proper motion -- except maybe some (or many ?) ejected run-away stars and
brown dwarfs as well as possibly ejected planets (or planetary mass objects). Hence, even if
GQ\,Lup\,A and the faint object show the same proper motion, they could be two gravitationally
independent members of the Lupus association, i.e. two young objects in the Lupus I cloud at
slightly different distances (even if only a few pc). This is possible for all apparently common
proper motion pairs, even in the field, but less unlikely in moving groups like T association and
star forming clouds.

\begin{figure}[ht]
\resizebox{\hsize}{!}{\includegraphics{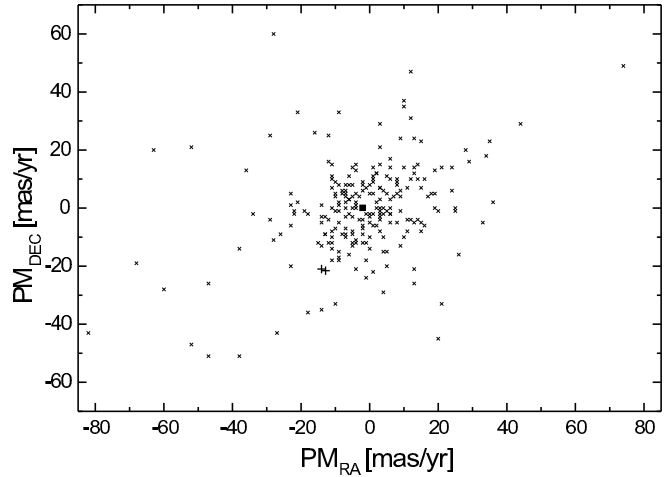}} \caption{Proper motions of stars listed in
Camargo et al. (2003) within 0.4 square degree around GQ\,Lup. GQ\,Lup and its companion are
plotted as black plus symbols and the average proper motion of all stars is shown as a black
square.}\label{pmcamargo}
\end{figure}

Let us now estimate the likelihood for this scenario or the significance by which we can reject
this possibility. There are 231 objects listed within 0.4 square degree around GQ\,Lup in Camargo
et al. (2003) with proper motions known (we use the Camargo et al. catalog here, because it has the
best precision in proper motions). The faintest detected objects in this catalogue exhibit a K -
band magnitude of 13.1\,mag, i.e. the brightness of GQ\,Lup\,b. Then, the probability to find one
such object within 0.7325\,$^{\prime \prime}$ around one other star is only 0.000075 which is still
regardless the (unlikely identical) proper motion.

According to our data in Table\,\ref{astrometry}, the difference in proper motion between star and
companion (candidate) is only $\rm{1.4\pm2.2}$\,mas/yr. None of the 231 stars within 0.4 square
degree around GQ\,Lup has the same proper motion as GQ\,Lup; there is not even any star within 3
times the given relative motion around the GQ\,Lup proper motion (Camargo et al. 2003), except
GQ\,Lup itself and its companion, see Fig.\,\ref{pmcamargo}. The total area population by those 231
stars around GQ\,Lup in the proper motions phase space is 17316\,$\rm{mas^{2}/yr^{2}}$. However
only a small fraction of stars exhibit high proper motions. The average proper motion of all star
is $\rm{PM_{RA}}$=-2\,mas/yr; $\rm{ PM_{DEC}}$=0\,mas/yr with a standard deviation of 17\,mas/yr,
assuming a normal distribution. The probability to find one object with a proper motion within
1.4\,mas/yr of GQ\,Lup\,A is then only 0.049, considering the non-homogeneous, but normal
distribution in Fig.\,\ref{pmcamargo}.

Hence, the probability to find one Lupus object with both almost the same total proper motion and
almost the same position of GQ\,Lup is only 0.000004, so that also this possibility can be rejected
by 4\,$\sigma$. And the probability to find one field L-dwarf (see above) with both almost the same
proper motion and almost the same position of GQ\,Lup is only $\rm{\le3\cdot10^{-10}}$, i.e.
negligible.

\subsection{The mean binary separation in Lupus}

In Nakajima et al. (1998), the surface distribution of companions as function of angular separation
is given for several different star forming regions including Lupus. From the power law fits in
their equation 14 for Lupus, one would expect 0.116 objects with a separation of
0.7325\,$^{\prime\prime}$ to the next star. Hence, the fact that we find one such object is
strongly deviant from the expectation (extrapolating their equation 14 for 1.0\,$^{\prime\prime}$
to 7.4\,$^{\prime\prime}$ slightly to 0.7325\,$^{\prime\prime}$).

Then, they specify the mean nearest neighbor distance (see their Fig.\,5) being 10 times the
so-called break point to be 7.4\,$^{\prime\prime}$ for Lupus, so that we get 1110\,AU as mean
nearest neighbor distance between any two objects in Lupus. In our case, we have a projected
physical separation of $\sim$\,100\,AU, again very much different.

Hence, we can conclude that the possibility of two objects with almost the same position and the
same proper motion, but slightly different distance within the Lupus I cloud, is very unlikely.

If there would be much more stars in Lupus I than known so far, e.g. strongly extincted within the
cloud, i.e. much more than taken into account above, then it could be less unlikely that the faint
object is not bound to GQ\,Lup\,A, but it would then become likely that it is bound to one of those
many other, as yet unknown, stars in Lupus I; hence, it would still be a companion.

If the very unlikely possibility of same proper motion, but slightly different distances by just a
few pc would be realized here, both objects would still be members of the Lupus I cloud, so that
the distance and age estimate for the faint object is still correct (as in N05a), and hence, also
the mass estimate. It would still be below the D burning mass limit, hence either a very low-mass
(sub-)brown dwarf or an ejected planet.

\subsection{Escape velocity in the GQ\,Lup and 2M1207 systems}

Common proper motion is not a proof for being gravitationally bound. The lower the total mass and
the larger the separation between the two co-moving companions, the lower the probability for being
gravitationally bound. We can calculate this as ejection velocity of the companion and as binding
potential of the system. To compare the recently presented system GQ\,Lup\, A and b with 2M1207 A
and b in these regards, we need the masses of all components and the physical separation between
them. According to the Wuchterl \& Tscharnuter (2003) model GQ\,Lup\,A is a
$\sim$\,$\rm{0.7}$\,$\rm{M_{\sun}}$ star; according to Hughes et al. (1994), when using the
unpublished Swenson et al. models or the D'Antona \& Mazzitelli (1994) models, the star may have
$\sim$\,0.5 to $0.3$\, $\rm{M_{\sun}}$, respectively, with an age of only $\sim$\,$10^{6}$ to
$\sim$\,$10^{5}$\,yr, respectively, but it is dubious whether those two models are applicable at
the young age of GQ\,Lup. The companion mass is negligible (we use 2\,$\rm{M_{Jup}}$). The
companion of the brown dwarf 2M1207 is separated by $\rm{0.772\pm0.004}$\,$^{\prime\prime}$, being
$\rm{54\pm16}$\,AU at $\rm{70\pm20}$\,pc as assumed in Chauvin et al. (2005), or just
$\rm{41\pm5}$\,AU at $\rm{53\pm6}$\,pc as given in Mamajek (2005). The mass of the host brown dwarf
A is assumed to be $\sim$\,25\,$\rm{M_{Jup}}$, that of the companion b to be
$\sim$\,5\,$\rm{M_{Jup}}$ (Chauvin et al. 2004), but both mass estimates are very uncertain due to
the low age and the fact that Chauvin et al. (2004) used only models which are not applicable at
young ages. Mamajek (2005) give slightly lower mass estimates, but according to the same models
which may not be applicable. For our calculations here, we use the values given in Chauvin et al.
(2004).

Both companions are co-moving with their primaries, but the common proper motion of such systems is
not a sufficient proof of companionship, because the escape velocity ($\rm{\sqrt{2GM/a}}$) for such
wide companions is much smaller than the proper motion of the system barycenter.

Due to the unknown inclination and eccentricity of the orbits, it is difficult to predict exact
escape velocities, so that we derive only rough estimates.

With the separation and the mass of the GQ\,Lup system, one would expect an escape velocity of
$\rm{5.2\pm2.1}$\,mas/yr for the companion of GQ\,Lup. By comparing the HST astrometry from 1999
with the NACO data from 2005, we derive a relative motion between GQ\,Lup and its companion of only
$\rm{1.4\pm2.2}$\,mas/yr, smaller than the expected escape velocity, but consistent with the
expected orbital motion ($\rm{v_{esc}/\sqrt{2}=3.7\pm1.5}$\,mas/yr). These results tend to confirm
that the common-proper motion pair GQ\,Lup is indeed bound. The measurement of the orbital motion
of the companion with accurate NACO astrometry (4\,mas) alone should be possible after a few more
years of epoch difference.

Chauvin et al. (2005) present relative astrometry for both components and we derive a relative
motion between both objects of $\rm{4.1\pm8.2}$\,mas/yr (using the data with the smallest
astrometric uncertainties taken in April 2004 and Feb 2005)\footnote{There seems to a typo in
Chauvin et al. (2005): The predicted motion $\rm{\Delta \delta}$ for the background hypothesis for
Feb 2005 in given as $\rm{-624\pm10}$\,mas in Table\,2 in Chauvin et al. (2005), but only
$\rm{-424\pm10}$\,mas according to our own calculation; all other numbers in that table seem to be
correct.}. However, the small total mass and relatively large separation of 2M1207 A and b yield an
escape velocity of the companion of only $\rm{2.7\pm0.9}$\,mas/yr. The measured relative motion
exceeds the escape velocity, but the uncertainty (8.2\,mas/yr) is too large to conclude whether the
companion of 2M1207 is indeed orbiting its host brown dwarf or is already disrupted (but still
showing similar proper motion). With an expected orbital motion of $\rm{1.9\pm0.6}$\,mas/yr, one
has to wait several more years to detected first indications of orbital motion with NACO, assuming
again 4\,mas astrometric accuracy.

\subsection{Long-term stability of GQ\,Lup and 2M1207 systems}

In the following, we will discuss the long-term stability of both systems. The binding potential
($\rm{-GM_{tot}/2a}$) is used as indicator of the system stability against external disturbances,
e.g. stellar encounters. We derive ($\rm{-3\pm1)\cdot10^6}$\,J/kg for GQ\,Lup, but only
($\rm{-2.4\pm0.7)\cdot10^5}$\,J/kg for 2M1207 at $\rm{70\pm20}$\,pc (or
($\rm{-3.1\pm0.4)\cdot10^5}$\,J/kg at $\rm{53\pm6}$\,pc), i.e. the companion of GQ\,Lup is bound 10
to 13 times more tightly than the presented companion of 2M1207. Because both systems are located
in star forming regions with increased stellar density, we expect that both systems will undergo or
already have underwent several stellar encounters. Due to its higher stability, it is much more
likely for the GQ\,Lup system to survive the remaining time in its star forming region than the
2M1207 system.

\begin{figure}[h]
\resizebox{\hsize}{!}{\includegraphics{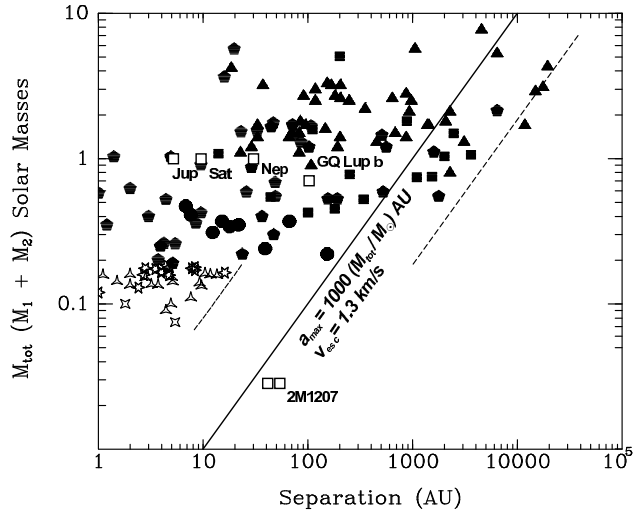}} \caption{The total mass of binary systems
versus their separation from Close et al. (2003) with very low-mass binaries as open stars and
normal stellar binaries as filled symbols. There are no low-mass systems with separations larger
16\,AU. The escape velocity of the low-mass systems is about 3\,km/s higher compared to more
massive wide binaries. The solid line fit $\rm{a_{max}=1000(M_{tot}/M_{\sun}}$)\,AU. Due to
stochastic encounters, binary systems with separations larger than that limit should evolve to
significantly wider orbits over time in the galactic disc, eventually being disrupted. The
companions of 2M1207 and GQ\,Lup are shown with open squares together with the giant planets in our
solar system. The GQ\,Lup system seems bound and long-term stable, the 2M1207 system not. The two
symbols plotted for 2M1207 are for the two different distances published, for 70\,pc (right)
according to Chauvin et al. (2005) and for 53\,pc (left) according to Mamajek (2005).}\label{sep}
\end{figure}

Even if both systems have survived their early time in the star forming region, they will undergo
many encounters with other stars passing by in the galactic plane. These events will perturb the
systems, slowly increasing their separations and eventually disrupting them. Following Weinberg et
al. (1987), Close et al. (2003) derive that only systems closer than
$\rm{a_{max}=1000(M_{tot}/M_{\sun}}$)\,AU are close enough not to significantly evolve, i.e. not to
disrupt in the galactic disc due to encounters with stars or clouds. In Fig.\,\ref{sep} we show
total masses of binaries versus their separation. The given long-term stability limit in the
galactic disk is illustrated as a solid line. Indeed most of the known binary systems are located
within this stability region, i.e. left of the solid line. For a total mass of 0.7\,$\rm{M_{\sun}}$
(GQ\,Lup), we find 700\,AU as maximal separation of long-term stable systems, but only 29\,AU for a
total mass of 30\,$\rm{M_{Jup}}$ (2M1207). With a separation of $\sim$\,54\,AU (or 41\,AU), the
2M1207 system exceeds this stability limit. It is therefore probable that it will be disrupted if
it is still bound. In contrast, the companion of GQ\,Lup fulfills the stability criteria, hence
this system should be long-term stable, like the giant planets in our solar system (see
Fig.\,\ref{sep}).

Close et al (2003) outline that very low-mass binaries are much tighter than their more massive
counterparts with escape velocities all being higher than 3.8\,km/s. They propose several scenarios
explaining the observed differences, e.g. ejection theories where brown dwarfs are formed because
they are ejected from their cluster by close encounters of other objects, being starved of
accretion material. Only those systems survive which are the tightest bound, being observable today
as common proper motion pairs. On the other hand brown dwarf binaries could also be the final
remnants of a cluster decay, a process which is also only survived by tightly bound systems. The
companion of 2M1207 is an exception of that rule because its escape velocity is only
$\rm{0.9\pm0.3}$\,km/s, i.e. this system falls below the limit proposed by Close et al. (2003).

\section{Summary and conclusions}

We have shown that for any of the 6 proper motions published for GQ\,Lup\,A, the proposed companion
(N05a) does indeed share its proper motion (by at least 6\,$\sigma$). For the weighted mean proper
motion, the common proper motion is very significant (6.9\,$\sigma$ + 3.9\,$\sigma$ +
7.8\,$\sigma$).

Even if taking into account the large error in distance or parallax ($\rm{140\pm50}$\,pc), the
significance remains large. The effect of this error on the significance is only
$\rm{\pm0.3}$\,dex.

Then, we also show that is it very unlikely that two objects with very similar proper motion are
found within 0.7\,arcsec on the sky. For the spectral type of the companion (L1-2), the probability
of chance alignment is below $3 \cdot 10^{-10}$. The mean nearest neighbor distance in Lupus is
1110\,AU (Nakajima et al. 1998), while we observe a projected physical separation of $\sim$\,100\,AU.

We also show that the observed change in separation between GQ\,Lup\,A and b is smaller than the
expected ejection velocity, so that the system is very likely bound. This is not yet shown for the
2M1207 system, where the observed change is separation may be larger than the expected ejection
velocity. Also, the binding potential of the GQ\,Lup system is much larger than for the 2M1207
system, so that the former system is long-term stable, the latter is not. 2M1207 is less stable,
even though of smaller separation, because of the lower masses of the components (as published in
Chauvin et al. 2004). However, if the component masses would be larger, than the system could be
bound and stable.

\acknowledgements We would like to thank Andreas Seifahrt for useful discussion about the proper
motion and astrometry of GQ\,Lup.

\end{document}